\documentclass[a4paper,11pt]{article}
\pdfoutput=1 
\usepackage{jinstpub} 
\usepackage{multirow}
\usepackage[above, below]{placeins}
\usepackage{acronym}
\usepackage{amsmath}
\usepackage{hyperref}
\usepackage{graphicx}
\usepackage{lineno}
\usepackage{siunitx}
\usepackage{subcaption}
\usepackage{textgreek}

\title{Timing performance of SiPM-on-Tile elements: Laboratory and test beam measurements}

\author[a,b,1]{F.~Hummer\note{Corresponding author.}}
\author[b]{L.~Emberger}
\author[a,b]{F.~Simon}
\affiliation[a]{Institute for Data Processing and Electronics, Karlsruhe Institute of Technology,\\ Hermann-von-Helmholtz-Platz 1, 76344 Eggenstein-Leopoldshafen, Germany}
\affiliation[b]{Max-Planck-Institut f\"ur Physik,\\Boltzmannstr. 8, 85748 Garching, Germany}

\emailAdd{fabian.hummer@kit.edu}

\abstract{The SiPM-on-Tile technology for highly granular calorimeters, where small plastic scintillator tiles are directly read out with SiPMs, has been developed for the CALICE Analog Hadron Calorimeter, and has been adopted for parts of the hadronic section of the CMS High Granularity Calorimeter. For future electron-positron colliders, a single cell time stamping on the sub-nanosecond level for energy deposits corresponding to single minimum-ionizing particles is desired to provide background rejection and to support pattern recognition and energy reconstruction with particle flow algorithms. To better understand the time resolution achievable with the SiPM-on-Tile technology, we have performed detailed measurements in test beams at DESY, probing scintillator tiles of different sizes. The study is complemented by laser measurements that provide insights into processes within the scintillator tile relevant for timing. In this publication, we will discuss our measurement methods and the results of our SiPM-on-Tile timing study.
}

\begin{document}
\maketitle
\flushbottom


\acrodef{ASIC}{application-specific integrated circuit}
\acrodef{AHCAL}{Analog Hadron Calorimeter}
\acrodef{EBU}{Ecal Base Unit}
\acrodef{FWHM}{full width at half maximum}
\acrodef{HBU}{Hcal Base Unit}
\acrodef{MC}{Monte Carlo}
\acrodef{MPPC}{Multi-Pixel Photon Counter}
\acrodef{RMS}{root mean squared}
\acrodef{SiPM}{Silicon Photomultiplier}
\acrodef{PMT}{Photomultiplier Tube}
\acrodef{LTP}{large technological prototype}
\acrodef{PFA}{particle-flow algorithms}

\acused{ASIC}
\acused{FWHM}
\acused{MPPC}
\acused{RMS}
\acused{PMT}


\newcommand{\tile}[1]{$#1\times#1\times3~\text{mm}^3$}
\newcommand{\cube}[1]{$#1\times#1\times#1~\text{mm}^3$}

\section{Introduction}

Timing in event and shower reconstruction has become one of the main directions in calorimeter development and shows promising potential for the improvement of  energy resolution in highly granular hadronic calorimeters \cite{Akchurin:2021afn, Graf:2022lwa, Chekanov:2022vyh, Benaglia:2262300}. High-precision time information enables the characterisation of the evolution of particle showers with the aim to identify shower components. Other uses of timing in calorimetry include background and pileup suppression, as well as the search for new long-lived particles. 

A Silicon Photomultiplier (SiPM) is a 2D array of avalanche photodiodes operated in parallel and read out via one electric connection. Their very small size, the high photon detection efficiency and fast response make SiPMs very attractive for high-energy physics experiments, resulting in wide-spread use \cite{SiPM_2_1811.03877v2}.

The SiPM-on-Tile technology has been developed for the CALICE Analog Hadron Calorimeter (AHCAL) \cite{CALICE:2022uwn}, and has been adopted for parts of the hadronic section of the CMS High Granularity Calorimeter (HGCAL) \cite{CERN-LHCC-2017-023}. The plastic scintillator tiles are square-shaped, wrapped in reflective foil, and are directly placed on a PCB which hosts the SiPMs and the front-end electronics. A central dimple in each tile houses the SiPM which reads out the light from that tile directly. This configuration is optimised for light collection and for automatic assembly with a glue dispenser and a pick-and-place machine.

To better understand the time resolution achievable with the SiPM-on-tile technology, we have performed detailed measurements in beam tests at the DESY II test beam facility \cite{DESY_TB_1807.09328v2}, probing different scintillator tile sizes. 
In addition, we introduce measurement methods that allow us to observe the time structure of the scintillation process, the time structure of the light collection in the scintillator tile, and the response of the SiPM and measurement electronics separately.

\subsection{Experimental Equipment}

\begin{figure}[h]
	\centering
	\includegraphics[width=0.9\linewidth]{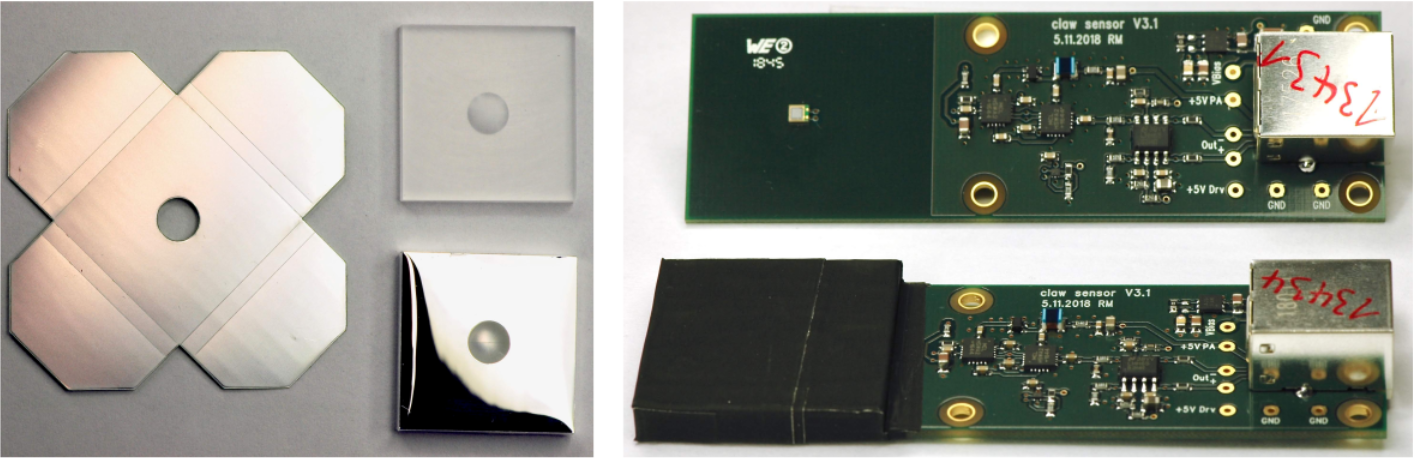}
	\caption{Left: An AHCAL-style scintillator tile of size \tile{30} with a dimple for the SiPM and reflective wrapping foil. Right: Two CLAWS boards, the upper one without scintillator, the lower one with a scintillator. The scintillator tile is fixed with black tape, which also shields from ambient light.}
	\label{claws-single-board}
\end{figure}

Modules based on a single SiPM-on-tile element were developed for the CALICE T3B experiment~\cite{T3B_Simon:1602997}. These modules were further developed and used within the CLAWS project for beam background studies and a beam abort system for the SuperKEKB collider~\cite{Gabriel:2020rqe, PhD_Miro, PhD_Hendrik}. 
For the measurements discussed in this publication, we used the latest iteration of the CLAWS boards, the phase 3 CLAWS shown in figure \ref{claws-single-board}.
Each boards contains a single SiPM under a plastic scintillator tile, as well as the corresponding amplifier electronics.

The SiPMs installed on the CLAWS boards are Hamamatsu S13360-1325PE~\cite{SiPM_S13360-1325PDE_Datasheet}, the same that are used in the AHCAL technological prototype \cite{CALICE:2022uwn}. On each CLAWS module, the signal from the SiPM is amplified by a BGA614 darlington transistor and converted to a differential signal using a LMH6553 differential amplifier.

Up to four CLAWS modules are connected to one receiver box using Cat6A ethernet cables. The cables are used both to supply power to the CLAWS modules and to transmit the differential analog signals from the sensor. The SiPM bias voltage for up to four CLAWS modules is generated in the receiver box and supplied to the front-end via the ethernet cables. LHM6553 differential amplifiers located in the receiver box convert the differential signals to 50~$\Omega$ single ended signals on BNC connectors. The full analog waveforms of each event are recorded with a PicoScope 6804E digital computer-controlled oscilloscope which has a bandwidth of 500~MHz \cite{PicoScope_Datasheet}.

The plastic scintillator tiles are packed in 3M Vikuiti Enhanced Specular Reflector (ESR) foil to increase the light collection efficiency. Black tape is used to mount the scintillator tile to the CLAWS board and as a stray light protection.

Unlike the scintillators in the AHCAL, which are made of injection-moulded polystyrene, for the present study general-purpose plastic scintillator BC408 \cite{BC408_Datasheet} is used, because this material is easily available and can be machined into different sizes.
We investigated the time resolution of three different scintillator tile sizes \tile{20}, \tile{30} and \tile{40}. Note that the dimension of the dimple is the same in all three cases. Drawings of the tile dimensions can be found in \cite{Malinda_and_Frank_Misalignment}.

An additional advantage of BC408 is that its properties are well known, which allows us to model the SiPM-on-tile setup in a Geant4 simulation. The results of the simulation will be discussed in an upcoming publication.

\subsection{Breaking down the Signal Creation}

In a SiPM-on-tile configuration, detecting a charged particle involves three subsequent steps:
\begin{enumerate}
    \item The particle deposits energy in the \textbf{scintillator} and photons are emitted.
    \item The photons propagate through the scintillator tile, and might be reflected at the foil or the tile surface. The term \textbf{light collection} summarises all the effects that influence the way of the photons from the scintillation spot to the SiPM.
    \item The photons that reach the SiPM can cause an avalanche in one of the pixels. The response of a single pixel is called photoelectron or p.e. The electrical signal is then amplified and measured with the oscilloscope.
\end{enumerate}

Note that all three of these processes potentially have an inherent time structure that influences the time resolution achievable with the SiPM-on-tile configuration. 

To understand these processes at a deeper level, multiple different measurement concepts are required. Naturally, the test beam measurements include all three steps of signal creation. Using a small scintillator cube without reflective foil, we can reduce geometric effects and estimate the emission times of the material without light collection. To eliminate the effects from scintillation, we can use pulsed laser light as an alternative light source.

All these measurements are however sensitive to step 3, the light detection and DAQ. Therefore, the impact of the SiPM and measurement electronics is measured separately by shooting short laser pulses directly at the SiPM and measuring the response.

\begin{figure}[h]
	\centering
	\includegraphics[height=4.8cm]{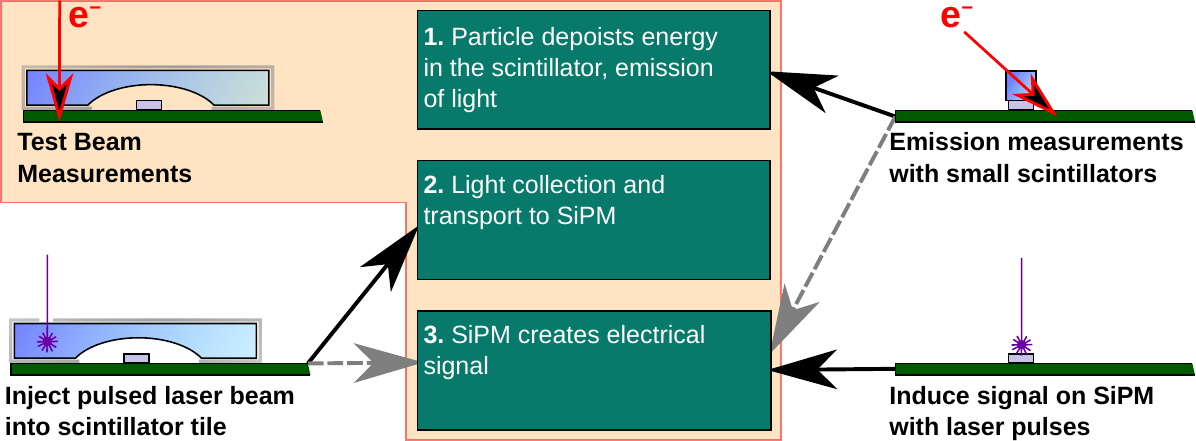}
	\caption{Illustration of the three steps that it takes to generate a signal in a SiPM-on-tile configuration, and the different measurement methods designed to investigate these steps separately.}
	\label{fig:signal-creation-steps}
\end{figure}

Figure \ref{fig:signal-creation-steps} illustrates the three signal creation steps as well as the different measurement methods. The following section \ref{sec:02-test-beam} shows our test beam setup and the measured time resolutions. Section \ref{sec:03-small-cubes} discusses the timing properties of BC408 measured with small scintillator cubes, and the laser measurements are presented in section \ref{sec:04-laser}. Section \ref{sec:discussion} contains a discussion of the photon time distributions measured with these different methods. 

\section{Test Beam Measurements}
\label{sec:02-test-beam}

Measurements with high-energy charged particles involve all three steps of signal creation (scintillation, light collection and light detection and measurement with the SiPM). We used electrons with an energy of 3~GeV at DESY's test beam facility \cite{DESY_TB_1807.09328v2}.
In two test beam weeks of the CALICE AHCAL group, in October 2020 and 2021, we placed our timing setup in front of the AHCAL prototype, such that both experiments used the same electron beam.

The two main observables of these measurements are the \textbf{light yield} and the \textbf{time resolution} of each SiPM-on-tile configuration. The light yield is defined as the most probable number of photoelectrons (p.e.) measured at the SiPM for a minimum ionising particle in the detector. The time resolution is derived from the  hit times reconstructed from individual waveforms recorded in the detectors, as discussed in further detail below.

\subsection{Experimental Setup}

\begin{figure}[ht]
  \center
    \includegraphics[width = 0.495\textwidth]{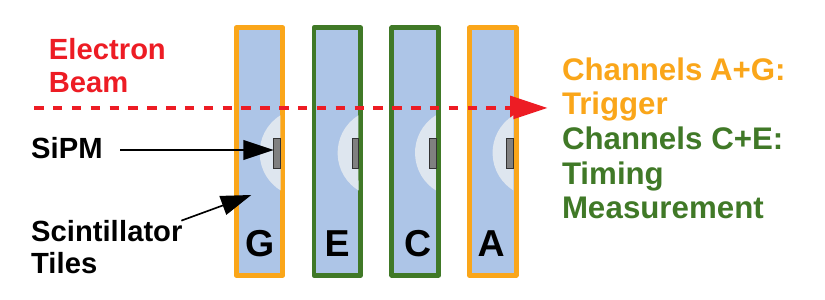}
    \caption{Arrangement of four scintillator tiles during beam test measurements for the SiPM-on-tile timing study}
    \label{fig:tb:setup-illustration}
\end{figure}

The timing setup that we used during the test beam measurements is a "beam telescope" made of four scintillator tiles as shown in figure \ref{fig:tb:setup-illustration}. The two outer scintillator tiles (channels A and G) are used for a coincidence trigger, and the two inner scintillator tiles (channels C and E) are used to calculate the time resolution.

In the first test beam campaign in October 2020 all four channels were recorded and the coincidence trigger was generated internally in the oscilloscope. The voltage resolution was 8~bit in a range from --1~V to +1~V and the trigger level was set to an equivalent of approximately 2.5 p.e.
For each event we recorded 500 pre-trigger and 1000 post-trigger samples, with a sampling time of 0.4~ns. This resulted in an overall time range of 600~ns for each waveform. 

In the second test beam campaign in October 2021 we additionally employed a Keysight P9242A oscilloscope to provide an external trigger for the PicoScope. This means that only the data channels C and E were recorded, at the benefit of a higher sampling time of 0.2~ns. In the analysis of this data set, an effective sampling time of 0.4~ns was used. Voltage and trigger settings were left unchanged with respect to the first test beam campaign.

Table \ref{tab:tb:runs} summarises the measurement settings of the data used in this publication.

The measurements are controlled by a python script that regularly (approximately every 5 minutes) interrupts for calibration. A calibration run is a separate measurement where each channel is recorded individually, one after another. The trigger level is set to approximately 0.5 p.e which means that it triggers mostly on dark counts where a single pixel of the SiPM fired. Accordingly, the voltage range is set to $\pm50$~mV. With these calibration measurements the stability of the SiPM gain is monitored for each channel. The integral of the calibration waveforms is used to estimate the number of photoelectrons for the recorded signal waveforms.

\begin{table}[h]
    \caption{Summary of measurement runs used to characterise the SiPM-on-tile with BC408 plastic scintillator tiles. The normal measurement runs, listed in the first three lines, are regularly interrupted for calibration runs where the 1 p.e. peaks of each channel are recorded separately.}
    \label{tab:tb:runs}
    \begin{tabular}{l|lllll}
        \textbf{Tile Size} & \textbf{Trigger Size} & \textbf{Voltage res.} & \textbf{Trigger} & \textbf{Date} \\
        \hline
        \tile{20} & \tile{20} & $\pm1$~V, 8~bit & External & October 2021 \\
        \tile{30} & \tile{30} & $\pm1$~V, 8~bit & Internal & October 2020 \\
        \tile{40} & \tile{30} & $\pm1$~V, 8~bit & External & October 2021 \\
        \hline
        Calibration & Trigger on 1 p.e. peak & $\pm50$~mV, 8~bit & Internal & \\
    \end{tabular}
\end{table}

\subsection{Results: Light Yield}
\label{sec:tb:ly}

The light yield of a detector module is defined as the most probable number of photoelectrons measured at the SiPM for minimum ionization (1 MIP of energy deposition) in the scintillator tile. The light yield is thus given by the most probable signal area of a minimum ionizing particle signal divided by the area of a  1~p.e. signal.
\begin{equation}
LY = \frac{area_\text{MIP}}{area_\text{1~p.e.}}
\end{equation}

\begin{figure}[ht]
  \center
  \begin{subfigure}{0.495\textwidth}
    \includegraphics[width = 1\textwidth]{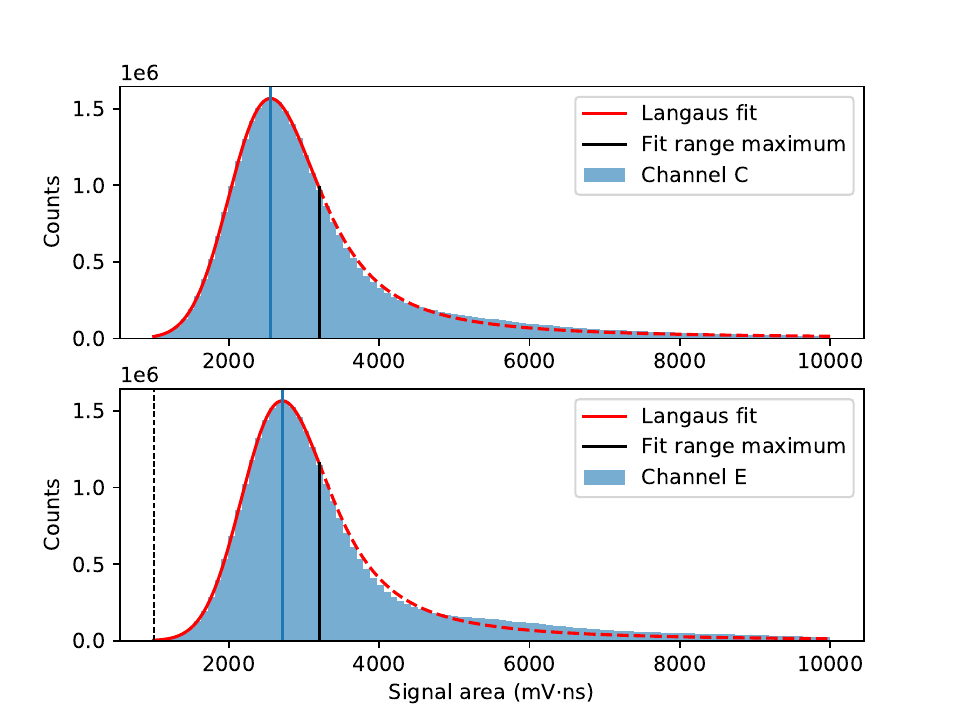}
    \caption{BC408 \tile{20}}
    \label{fig:tb:langauss-20x20}
  \end{subfigure}
  \begin{subfigure}{0.495\textwidth}
    \includegraphics[width = 1\textwidth]{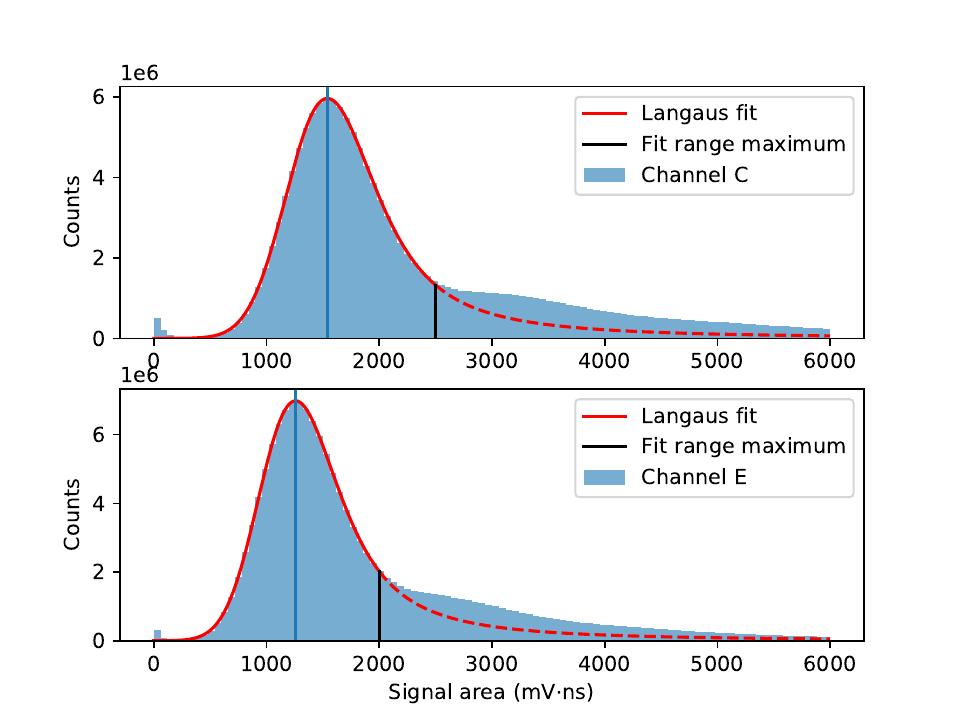}
    \caption{BC408 \tile{30}}
    \label{fig:tb:langauss-30x30}
  \end{subfigure}
  \begin{subfigure}{0.495\textwidth}
    \includegraphics[width = 1\textwidth]{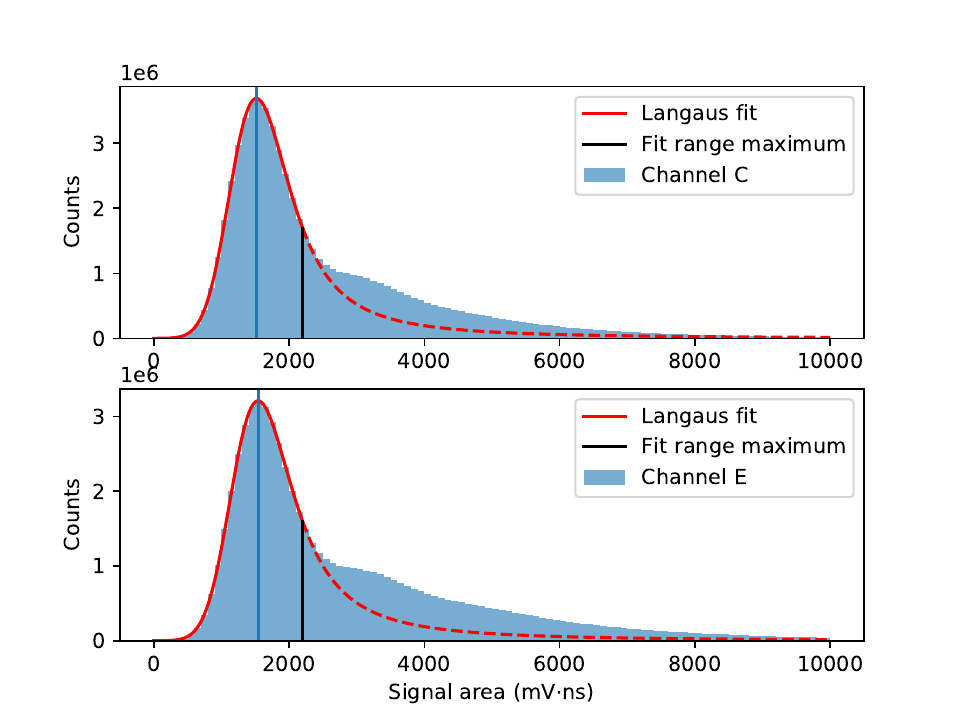}
    \caption{BC408 \tile{40}}
    \label{fig:tb:langauss-40x40}
  \end{subfigure}
  \begin{subfigure}{0.495\textwidth}
    \includegraphics[width = 1\textwidth]{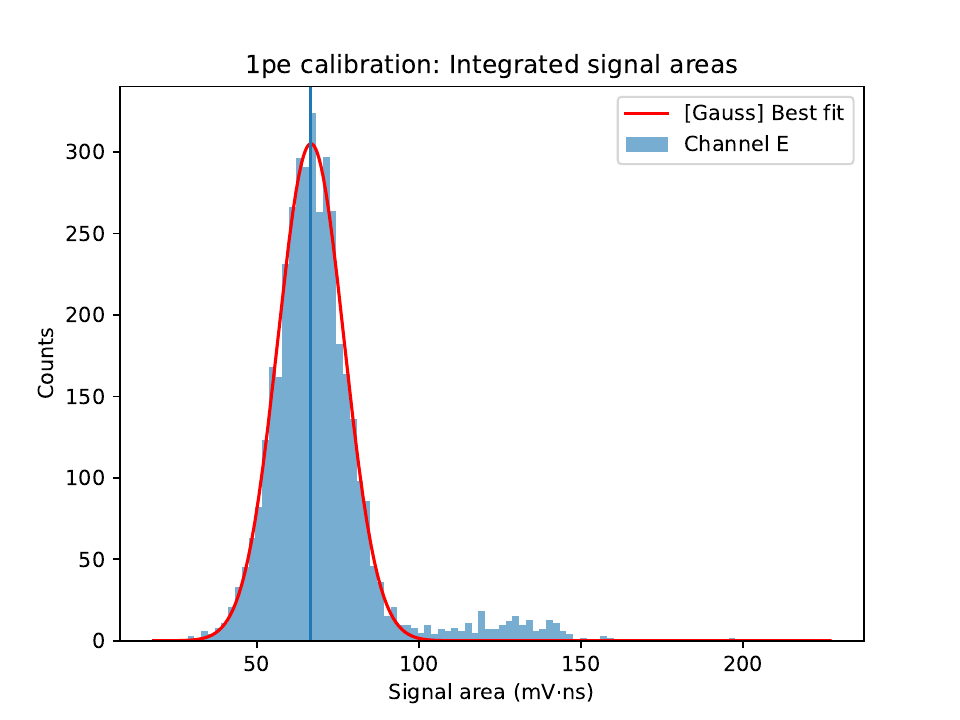}
    \caption{Calibration measurement: 1 p.e. peaks}
    \label{fig:tb:gaussian-1pe}
  \end{subfigure}
  \caption{Distributions of the area under the waveform for each channel of the test beam measurements and for a 1 p.e. calibration measurement. A numerical fit with a Landau/Gauss convolution (subfigures a-c), and a Gaussian in (subfigure d) is performed for each distribution. The black vertical lines indicate the maximum of the fit range. The best fit functions is plotted in red.}
  \label{fig:tb:langauss-all}
\end{figure}

The area of a 1 p.e. signal is extracted from the calibration measurement of each channel using a Gaussian fit as shown in figure \ref{fig:tb:gaussian-1pe}. There are a few counts at the double signal area, which arise from cross-talk or after pulses following a thermally excited 1 p.e. pulse.

Figure \ref{fig:tb:langauss-20x20}-\ref{fig:tb:langauss-40x40} shows the distributions of the signal areas, corresponding to the energy deposition in the scintillator tiles, for charged particles at the test beam. The hit energy distribution is skewed towards higher energies, and is expected to follow the characteristic behaviour of a Landau distribution.

To extract the most probable energy deposition, we performed a fit to a Landau distribution convolved with a Gaussian distribution, commonly referred to as "Langauss". The Gaussian part accounts for detector noise and measurement uncertainties.
The energy deposition in the smallest \tile{20} scintillator tiles (figure \ref{fig:tb:langauss-20x20}) agrees very well with the expected Langauss behaviour.

The distributions for the larger scintillator tiles (figures \ref{fig:tb:langauss-30x30} and \ref{fig:tb:langauss-40x40}) on the other hand show an excess of events at energies above the minimum ionization peak\footnote{Especially the run with the \tile{40} scintillator tiles (figure \ref{fig:tb:langauss-40x40}) has a high number of multi-particle events, for once because larger tiles have a higher geometric cross section, and on the other hand because a beam collimator with a larger aperture was used during this measurement.}.
This can be attributed to events where more than one particle deposits energy in the scintillator tiles. To make sure that no double particle events are included in the analysis, an upper limit for the fit range was set manually before the fit, indicated by the black lines in figure \ref{fig:tb:langauss-all}. Effectively, these limits correspond to an energy of approximately 1.5~MIP.

The additional particles can be either additional primary electrons from the test beam, or bremsstrahlung photons. 
A detailed discussion of the double particle events can be found in \cite{MSc_Fabian}.

Table \ref{tab:tb:ly} shows the resulting light yields in units of p.e./MIP. Manufacturing imperfections and individual fluctuations lead to a variation in the light yield for different detector modules. However, the variation between the data channels is consistent with the light yield variation of 12\% (standard deviation) observed in more than 20000 channels of the AHCAL technological prototype \cite{CALICE:2022uwn}.

Previous experimental studies show that the light yield scales as $1/\sqrt{A}$, where A is the area of the scintillator tile \cite{Malinda_and_Frank_Misalignment, A_vs_LY_FermiLab_TB}. This is consistent with our measurements.

\begin{table}[h!]
  \caption{Light yields in photoelectrons per MIP of the BC408 plastic scintillator tiles measured with the CLAWS boards at the test beam.}
  \centering
  \begin{tabular}{l|cc}
    Tile size & LY Ch. C (p.e./MIP) & LY Ch. E (p.e./MIP)\\
    \hline
    \tile{20} & 33.43 $\pm$ 0.13 & 37.67 $\pm$ 0.20 \\
    \tile{30} & 23.54 $\pm$ 0.17 & 19.80 $\pm$ 0.14 \\
    \tile{40} & 19.30 $\pm$ 0.12 & 19.42 $\pm$ 0.14 \\
  \end{tabular}
  \label{tab:tb:ly}
\end{table}

\subsection{Results: Time Resolution}
\label{sec:tb:time-res}

The first step in calculating the time resolution is to assign a particle hit time to each recorded event on each tile. 
We decided to use the amplitude-dependent constant fraction method: The point in time where a waveform first crosses 25\% of its maximum amplitude is used as the hit time. 

\begin{figure}[ht]
    \includegraphics[width = 1\textwidth]{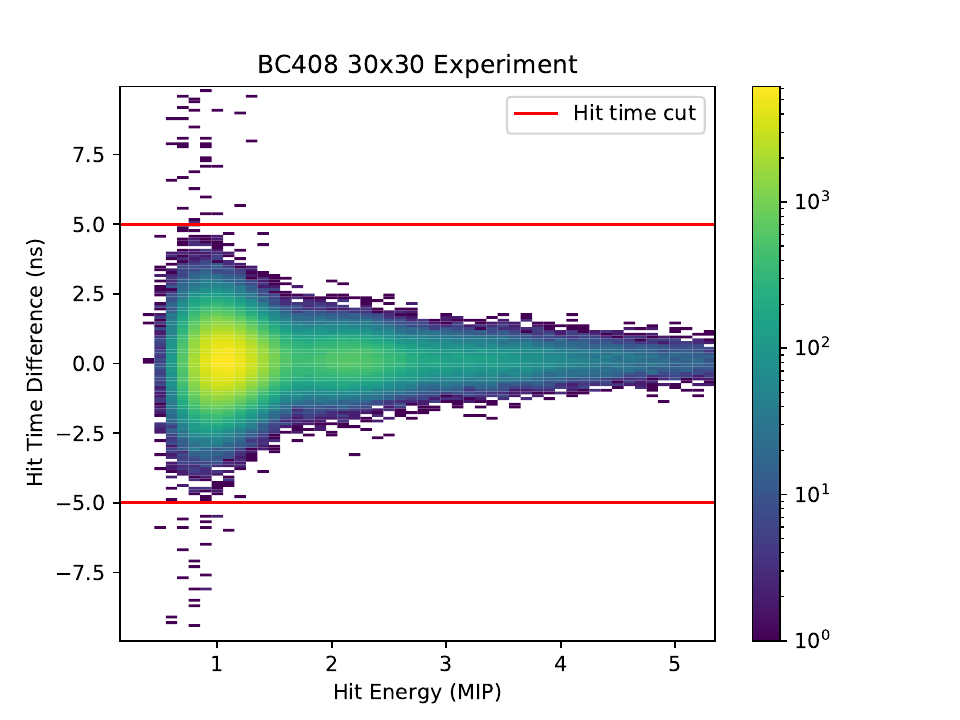}
    \caption{Histogram of hit time difference between the measurement channels and energy deposition in the scintillator for two BC408 \tile{30} scintillator tiles. Hits outside of the red lines are early hits that arise from noise and are rejected in the analysis. Only events with a similar energy deposition in both scintillator tiles are used in this plot.}
    \label{fig:tb:hit-time-diff}
\end{figure}

To eliminate trigger effects, we calculate the hit time difference between the two channels C and E.
Figure \ref{fig:tb:hit-time-diff} shows the correlation between the energy deposition in the scintillator tile and the hit time difference. For this analysis we only use events where the energy deposition is in the same bin in both channels. For each energy bin, we extract the width $\sigma$ of the hit time distribution with a Gaussian fit.

Hit times outside of the interval $\pm5~$ns are rejected since they tend to be due to early hits caused by baseline noise on one of the two channels.

Assuming that the hit times of channels C and E are uncorrelated, the time resolution of a single channel can be obtained by dividing the width $\sigma$ by $\sqrt{2}$. The following calculation is performed separately for each energy bin:

\begin{equation}
    \sigma_t = \frac{\sigma(t_C - t_E)}{\sqrt{2}}
\end{equation}

In this study, we assume that the "stochastic term" $\propto 1/\sqrt{E}$, resulting from the statistical nature of photon counting, is the only relevant contribution to the energy-dependent time resolution\footnote{Uncertainties often have additional terms that are constant or scale with $1/E$, but if we use a fit function that includes these terms, they are very small or even consistent with zero for our measurement results. It is thus sufficient to discuss the time resolution based only on the stochastic term.}.

\begin{equation}
    \sigma_t(E) = \frac{\sigma_{1MIP}}{\sqrt{E}}
    \label{eq:energy-binned-tr}
\end{equation}

\begin{figure}[ht]
    \includegraphics[width = 1\textwidth]{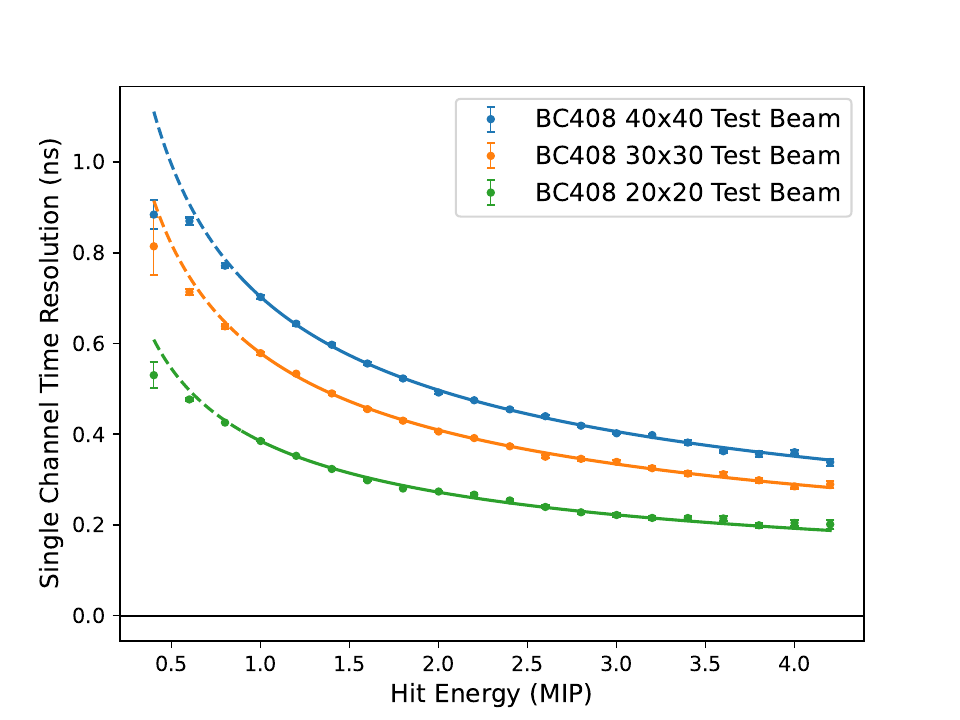}
    \caption{Time resolution of BC408 plastic scintillator tiles as a function of deposited energy in the scintillator. The data points are results of the test beam measurements, the lines are fits to a function following equation \ref{eq:energy-binned-tr}. Only data points at 1 MIP and above are included in the fit. The dashed lines show an extrapolation of the fit function to lower energies.}
    \label{fig:tb:tr-vs-energy}
\end{figure}

Figure \ref{fig:tb:tr-vs-energy} shows the measured time resolutions for all three scintillator tile sizes as a function of the energy deposition. The solid lines are fits to functions following equation \ref{eq:energy-binned-tr}. The only fit parameter, $\sigma_{1MIP}$, is the time resolution for a signal amplitude of 1 MIP. Energy depositions below 1 MIP are excluded from the fit due to the larger uncertainties of those measurement points. Table \ref{tab:tb:tr} shows the fit results for $\sigma_{1MIP}$. 

Both the time resolution and the light yield clearly depend on the scintillator tile size.
Since a higher number of photons arriving at the SiPM leads to a better time resolution (as seen in equation \ref{eq:energy-binned-tr}), the time resolution also improves with a higher light yield. This can be part of an explanation why the smaller scintillator tiles have a better time resolution.

 The fit results are only valid for one individual pair of scintillator tiles. The expected tile-to-tile variation of 12~\% of the light yield results in tile-to-tile fluctuation in time resolution of 6\%, if we consider the observed scaling of the time resolution with the square root of the light yield. The right column of table \ref{tab:tb:tr} shows the total uncertainty of the time resolution, including tile-to-tile fluctuations.

\begin{table}[h!]
  \caption{Single channel time resolution at minimum ionization $\sigma_{1MIP}$ for different BC408 scintillator tiles measured at the test beam. The total uncertainty includes the uncertainty from the fit as well as expected fluctuations in the time resolution that arise from tile-to-tile differences. The rightmost column is a comparison of the time resolution scaling with the tile area, assuming that the light yield is the only relevant effect.}
  \centering
  \begin{tabular}{l|cccc|c}
    & $\sigma_{1MIP}$ & Fit & Fit uncertainty & Total uncertainty & $\sigma_t \cdot A^{-0.25}$ \\
    Tile size & (ns) & red. $\chi^2$ & (ns) & (ns) & $\text{ns}/\sqrt{\text{cm}}$ \\
    \hline
    \tile{20} & 0.3847 & 3.10 & 0.0004 & 0.023 & $0.272\pm0.016$ \\
    \tile{30} & 0.5785 & 1.54 & 0.0006 & 0.035 & $0.33\pm0.02$ \\
    \tile{40} & 0.7027 & 1.76 & 0.0008 & 0.042 & $0.35\pm0.02$ \\
  \end{tabular}
  \label{tab:tb:tr}
\end{table}

As a consequence of equation \ref{eq:energy-binned-tr}, the time resolution of each scintillator tile scales with $\sigma_t \propto 1/\sqrt{LY}$. In addition, we know that the light yield scales with $LY \propto 1/\sqrt{A}$. If the light yield of the scintillator tile was the only contribution to the time resolution, then we would expect that $\sigma_t \cdot A^{-0.25}$ is constant. However, as shown in table \ref{tab:tb:tr}, this quantity deviates by more than $3\sigma$ between \tile{20} and \tile{30}. This indicates that there is an additional geometry-dependent contribution to the time resolution, which will be investigated in the following.

\section{Timing Properties of BC408}
\label{sec:03-small-cubes}

\begin{figure}[ht]
  \center
  \begin{subfigure}{0.39\textwidth}
    \includegraphics[height=7cm]{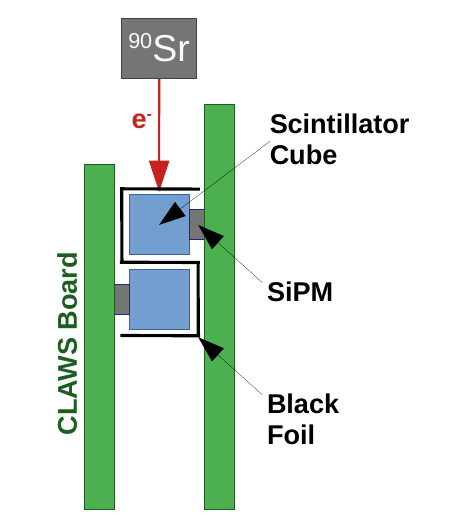}
    \caption{Illustration of the measurement setup.}
    \label{fig:cube:setup-illustration}
  \end{subfigure}
  \qquad
  \begin{subfigure}{0.48\textwidth}
    \includegraphics[height=7cm]{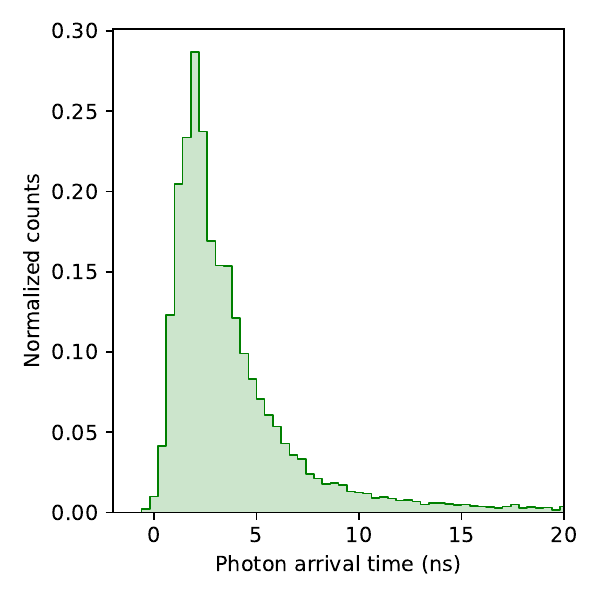}
    \caption{Measured emission time spectrum of BC408}
    \label{fig:cube:BC408-emission}
  \end{subfigure}
  \caption{Small scintillator cubes are used to mitigate geometric effects when measuring the photon emission time spectrum of the scintillator material. \textbf{Left:} An illustration of the setup. The scintillator cubes are made of BC408 and have a dimension of \cube{5}. A black foil shields the cubes from ambient stray light and crosstalk. The oscilloscope is configured as a coincidence trigger on both channels. \textbf{Right:} Photon time spectrum extracted from the measured waveforms using decomposition of the waveform into 1 p.e. peaks.}
\end{figure}

For measuring the timing properties of BC408, we use a setup that is designed to avoid geometric effects as far as possible and only detects light directly emitted from the scintillator. 
Scintillator cubes measuring \cube{5} are used instead of larger tiles to minimize the path that photons travel. Instead of ESR foil, the cubes are wrapped in black foil to suppress multiple reflections.
The scintillator cubes are excited using electrons from a $^{90}$Sr source. The signal of one cube is used for the timing analysis, and the second cube is used to provide a coincidence of two cubes as a trigger signal. This ensures full penetration of the first cube, and thus scintillation light emission over the full depth of the scintillator element.  The measurement setup is illustrated in figure \ref{fig:cube:setup-illustration}, and the measurement settings are summarised in table \ref{tab:measurement-settings}.

Since there is only one measurement channel, it is not possible to calculate the time resolution from a hit time difference. Instead we analyze the time structure of each waveform. 

Each photon triggers the same pixel response in the SiPM and the shape of this 1 p.e. signal is known from calibration measurements. Therefore, the arrival times of single photons at the SiPM can be estimated from the analog waveform measured with the oscilloscope. 
To this end we have developed a waveform decomposition algorithm, inspired by the signal reconstruction in the T3B experiment \cite{T3B_Simon:1602997}. Starting from the measured waveform, the algorithm iteratively subtracts 1 p.e. signals until only the baseline noise remains. The points in time where the 1 p.e. responses were subtracted are used as photon hit times.

For each event, the photon hit times are corrected by the time of arrival of the particle, which is determined using a 25~\% constant fraction on the raw signal. This step is necessary to mitigate the effects of trigger jitter. 

A detailed description of the waveform decomposition algorithm, as well as a discussion of its performance, can be found in \cite{MSc_Fabian}.

The photon emission time distribution of BC408 resulting from the waveform decomposition is shown in figure \ref{fig:cube:BC408-emission}. This distribution, which is a material property of BC408, will be discussed together with the light collection process in section \ref{sec:discussion}.

\begin{table}[h]
    \caption{Measurement settings used for determining the timing properties of BC408, and for the laser measurements.}
    \label{tab:measurement-settings}
    \centering
    \begin{tabular}{l|lll}
        \textbf{Measurement} & \textbf{Voltage res.} & \textbf{Sampling} & \textbf{Trigger} \\
        \hline
        BC408 \cube{5} cubes & $\pm500$~mV, 8~bit & 0.4~ns & Coincidence of both cubes \\
        Laser measurements & $\pm1$~V, 8~bit & 0.4~ns & Laser driver\\
    \end{tabular}
\end{table}

\section{Laser Measurements}
\label{sec:04-laser}

In addition to scintillation, there are two more processes involved in the signal creation within a SiPM-on-tile setup: light collection and light detection. Light collection means the propagation of scintillation photons through the tile until their absorption by the SiPM. Light detection includes the avalanches in the SiPM and the amplification and measurement of the resulting electrical signal.
These two processes can be probed individually by using a pulsed laser as an alternative light source.

For the laser measurements we used a PicoQuant LDH-P-C-440B pulsed diode laser. Detailed information on the setup can be found in \cite{MSc_Fabian}. The wavelength of 440~nm is close to the SiPM's peak sensitivity and similar to the peak emission wavelength of the used scintillator BC408. The pulses are shorter than 80~ps, which is significantly faster than the typical signals that we measure for example at the test beam. The measurement settings are summarised in table \ref{tab:measurement-settings}.

\begin{figure}[ht]
  \center
  \begin{subfigure}{1\textwidth}
    \includegraphics[width = 1\textwidth]{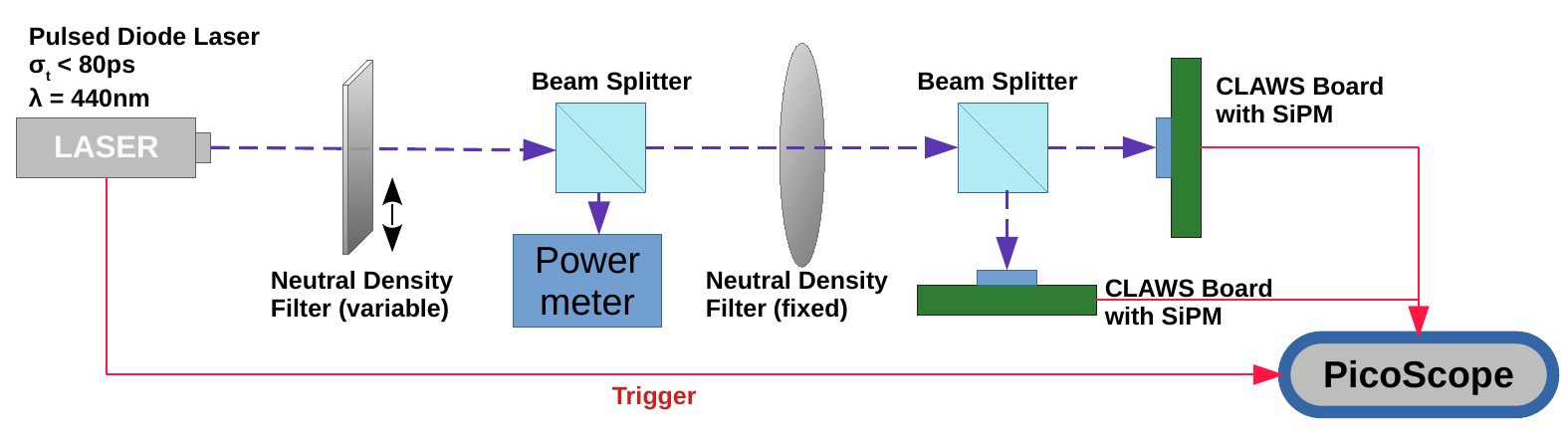}
    \caption{Sketch of the setup}
  \end{subfigure}
  \begin{subfigure}{1\textwidth}
    \includegraphics[width = 1\textwidth]{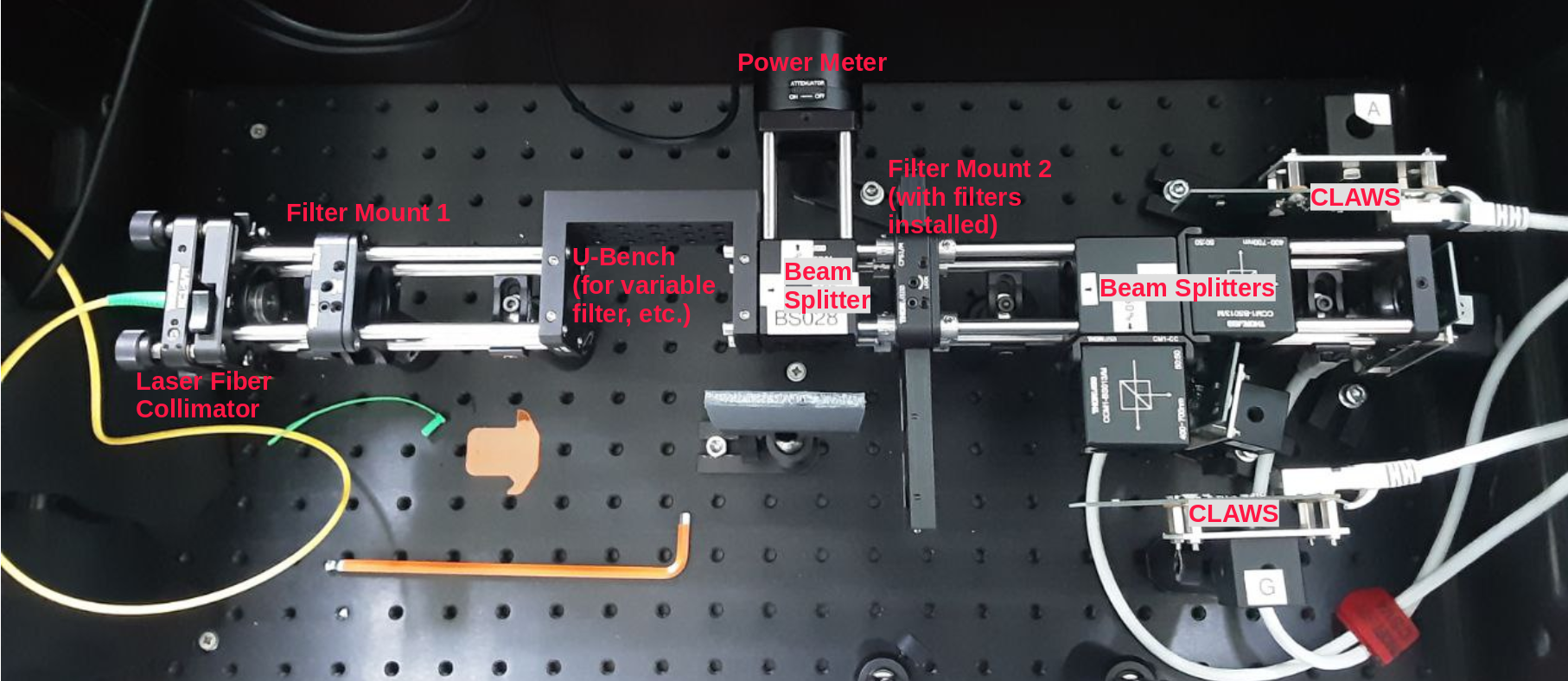}
    \caption{Picture of the setup.}
  \end{subfigure}
  \caption{Laser setup used to study light collection and light detection in the SiPM-on-tile setup without the impact of scintillation light. Short laser pulses of variable intensity are directed to multiple SiPMs on CLAWS modules simultaneously.}
  \label{fig:laser:setup}
\end{figure}

Figure \ref{fig:laser:setup} shows an illustration and a photograph of the laser setup. The light intensity that reaches the SiPMs is controlled using a set of neutral density optical filters and is monitored with a power meter. The beam is split and directed to two different SiPMs, which allows us to calculate the time resolution from the hit time difference, as described in section \ref{sec:tb:time-res}. The trigger signal is directly provided from the laser driver.

To probe the response of the SiPM and electronics to very short light pulses, the laser beam is aimed directly onto the SiPM. This allows us to estimate the timing uncertainty that arises from the light sensor, the electronics and the data acquisition, i.e. the timing uncertainty of the light detection.

The light collection is probed by injecting short light pulses into the scintillator tile. Since the laser wavelength is close to the emission wavelength of the scintillator, there should be no more fluorescence such that this measurement is only sensitive to the geometric effects of light collection.

\section{Discussion of Photon Time Distributions}
\label{sec:discussion}

\subsection{Light Detection and Hardware Response}

\begin{figure}[ht]
  \center
  \begin{subfigure}{0.47\textwidth}
    \includegraphics[width = 1\textwidth]{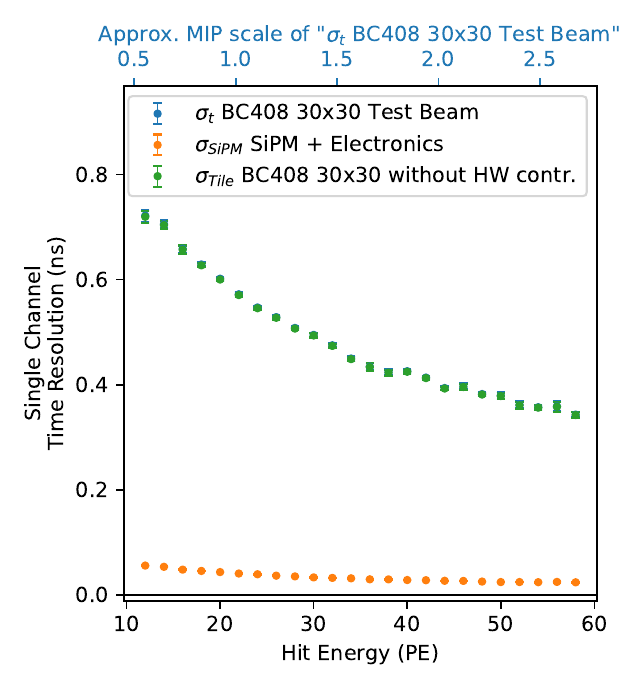}
    \caption{Energy dependent time resolution}
    \label{fig:laser:result-hardware}
  \end{subfigure}
  \begin{subfigure}{0.47\textwidth}
    \includegraphics[width = 1\textwidth]{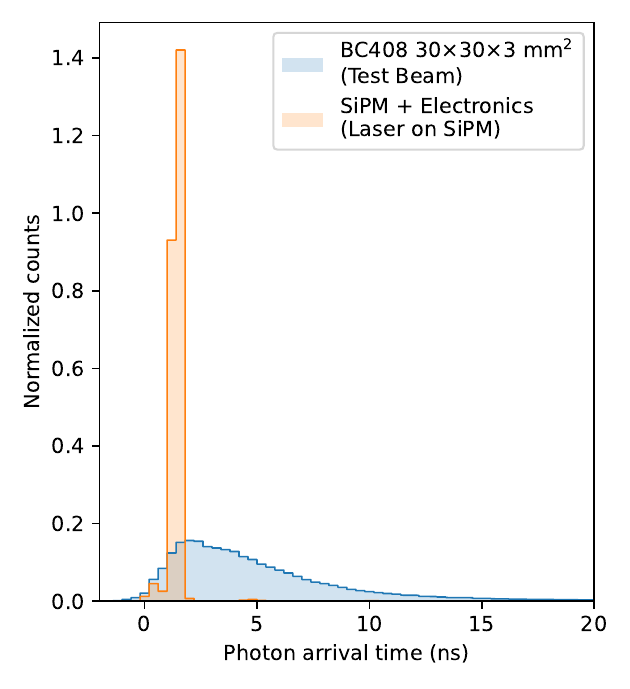}
    \caption{Photoelectron time distribution}
    \label{fig:laser:result-distr}
  \end{subfigure}
  \caption{Comparison of the test beam measurements (blue) and a pulsed laser signal directly on the SiPM (orange). The \textbf{left plot} compares these two measurements in terms of energy-dependent time resolution. The green curve is the computed time resolution of the test beam measurement without the timing uncertainties of the SiPM. The common energy scale is photoelectrons, and for reference the MIP scale of the test beam measurement is shown at the top axis. The \textbf{right plot} compares the photoelectron time distribution of test beam and laser-on-SiPM measurements. The photon hit times were extracted by decomposition of the waveform into 1 p.e. peaks (see section \ref{sec:03-small-cubes}).}
\end{figure}

Figure \ref{fig:laser:result-hardware} shows the time resolution of the light detection which was obtained from laser measurements, compared to the time resolution of a \tile{30} scintillator tile at the test beam. If we assume that the timing uncertainty from the electronics and SiPM is uncorrelated to the effects of the scintillator tile (scintillation and light collection), we can write the overall time resolution of the SiPM-on-tile system as
\begin{equation}
    \sigma_t = \sigma_{Tile} \oplus \sigma_{SiPM} = \sqrt{\sigma_{Tile}^2 + \sigma_{SiPM}^2}
\end{equation}

The green data points in figure \ref{fig:laser:result-hardware} are the time resolution of the test beam measurement \textit{without} the uncertainties induced by the time resolution of the SiPM:
\begin{equation}
\sigma_{Tile} = \sigma_{t} \ominus \sigma_{SiPM} = \sqrt{\sigma_{t}^2 - \sigma_{SiPM}^2}
\end{equation}

Since the timing uncertainty of electronics and SiPM is almost an order of magnitude lower than the overall time resolution, and since uncertainties are added in quadrature, this impact is negligible in our measurements.

This can also be seen the time distribution of photons arriving at the SiPM, shown in figure \ref{fig:laser:result-distr}. Compared to the time distribution of scintillation photons, the laser measurement yields a very short time distribution. Further discussion on this plot can be found in \cite{MSc_Fabian}.

All in all, the laser measurements show that the SiPM, CLAWS, amplifier electronics and oscilloscope together are still significantly faster than other signal components. Therefore, the contribution of the hardware to the time structure of the measured signals can be neglected.

\subsection{Scintillation and Light Collection}
\label{sec:discussion-conv}

\begin{figure}[ht]
    \center
    \begin{subfigure}{0.32\textwidth}
        \includegraphics[width = 1\textwidth]{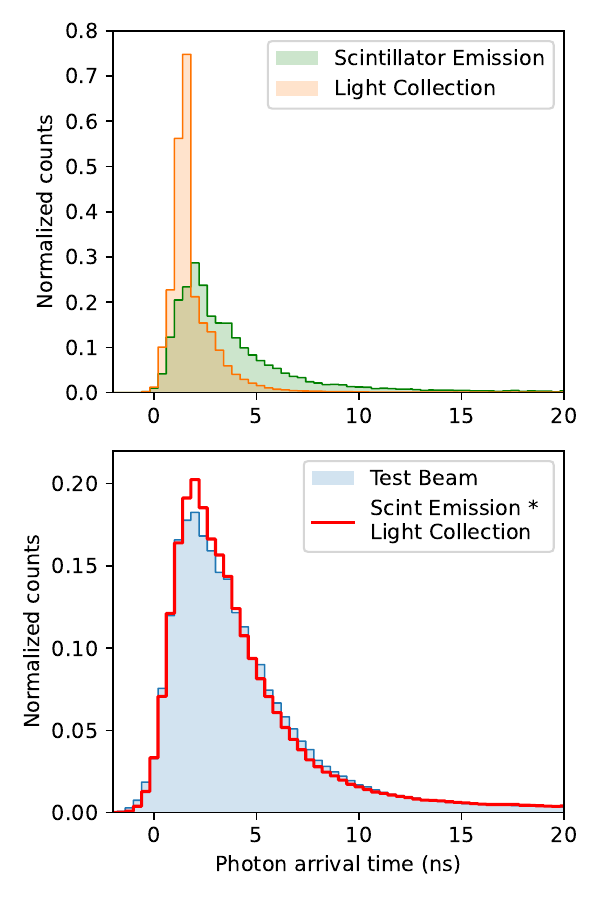}
        \caption{\tile{20}}
    \end{subfigure}
    \begin{subfigure}{0.32\textwidth}
        \includegraphics[width = 1\textwidth]{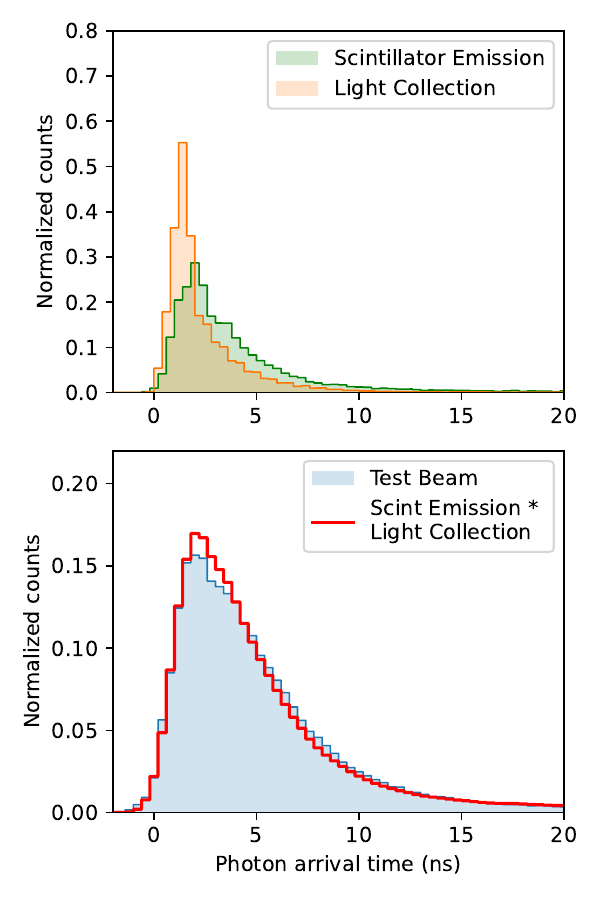}
        \caption{\tile{30}}
    \end{subfigure}
    \begin{subfigure}{0.32\textwidth}
        \includegraphics[width = 1\textwidth]{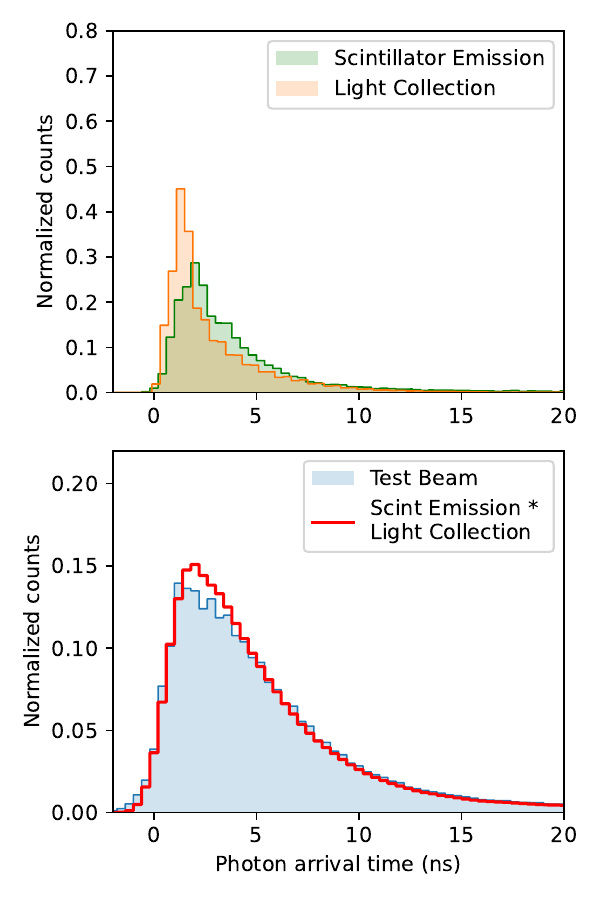}
        \caption{\tile{40}}
    \end{subfigure}
    \caption{Top row: The photoelectron time distribution of the light collection process for BC408 scintillator tiles of different sizes is shown in orange. The green histogram is the photon emission time distribution of BC408 which is also shown in figure \ref{fig:cube:BC408-emission}. Bottom row: Photoelectron time spectrum observed at the test beam with the same tile geometry, and a convolution of scintillation and light collection. The photon hit times were extracted by decomposition of the waveforms into 1 p.e. peaks (see section \ref{sec:03-small-cubes}).}
    \label{fig:laser:lc-and-scint}
\end{figure}

Since the electronics and light detection do not contribute significantly to the overall timing properties of the signals that we observe at the test beam, the two remaining processes are light collection and scintillation. The upper panels of figure \ref{fig:laser:lc-and-scint} show the photon time distributions of these two processes for BC408 scintillator tiles: The light collection was measured with the laser setup and the scintillation histogram is the same as in figure \ref{fig:cube:BC408-emission}.

The lower panels of figure \ref{fig:laser:lc-and-scint} show the photon time distribution for BC408 scintillator tiles measured at the test beam. The time structures of the test beam signals agree very well with the red lines, which are convolutions of the scintillation and light collection distributions from the upper plots. This demonstrates that the signals that we observe at the test beam are indeed determined by scintillation and light collection together.

\subsection{Quantitative Discussion of Photon Time Distributions}
\label{sec:discussion-time-distr}

For a quantitative discussion we calculate the rise time, fall time and FWHM of the histograms shown in figure \ref{fig:laser:lc-and-scint} using a Monte Carlo method. The following procedure is applied to each of the histograms:
\begin{itemize}
    \item 10000 random photon hits are generated by sampling from the histogram.
    \item The randomly generated photons are again binned into a histogram with a bin of width 0.4~ns, and a univariate spline is calculated as an interpolation function. 
    \item The rise time, fall time and FWHM are calculated from the univariate spline function. The rise time and fall time are defined as the time between 20~\% and 80~\% of the maximum signal amplitude. 
    \item The above process is repeated 1000 times, and a mean and standard deviation for rise time, fall time and FWHM are calculated.
\end{itemize}

\begin{figure}[ht]
  \center
    \begin{subfigure}{0.32\textwidth}
        \includegraphics[width = 1\textwidth]{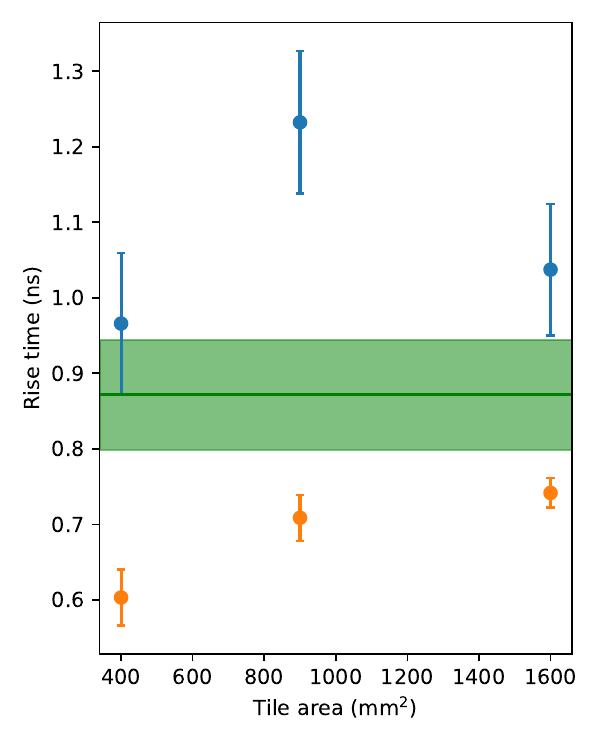}
        \caption{Rise Time}
    \end{subfigure}
    \begin{subfigure}{0.32\textwidth}
        \includegraphics[width = 1\textwidth]{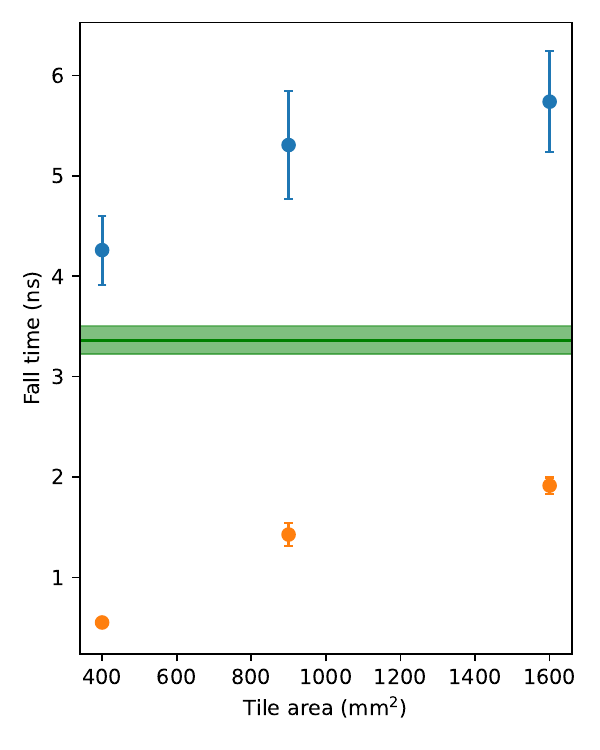}
        \caption{Fall Time}
    \end{subfigure}
    \begin{subfigure}{0.32\textwidth}
        \includegraphics[width = 1\textwidth]{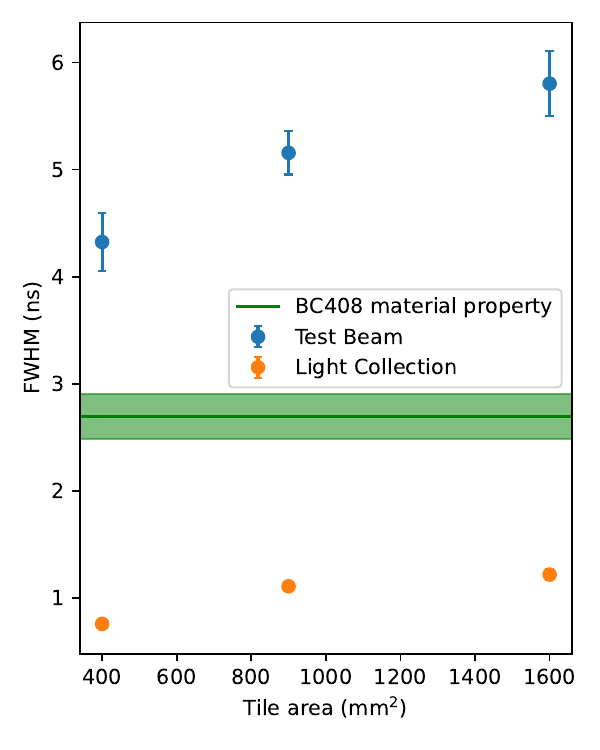}
        \caption{FWHM}
    \end{subfigure}
    \caption{Rise time, fall time and FWHM of photon time distributions for test beam measurements, scintillation and light collection. The values of all data points are listed in table \ref{tab:results-rise-and-fall-times}.}
    \label{fig:laser:lc-time-structure}
\end{figure}

The calculated statistics for the photon time distributions are visualized in figure \ref{fig:laser:lc-time-structure}. 

The rise time of the light collection process shows a slight increase with tile size which indicates that in a smaller scintillator tile there is a smaller time spread of the first photons that reach the SiPM. However, the rise times of the overall process observed at the beam test show no clear trend for different tile sizes. 

The fall times and FWHMs of the photon time distributions measured at the beam test show a clear dependence on the scintillator tile size. This means that larger scintillator tiles generally produce a wider photon time distribution, which can also be seen in figure \ref{fig:laser:lc-and-scint}. A possible explanation is that the photons have a longer mean free path in a larger scintillator tile which increases the time over which they can reach the SiPM. This would mean that the loss of photons is not dominated by absorption in the scintillator material\footnote{According to the manufacturer, BC408 has an absorption length of 2.1m and a refractive index of 1.58. Therefore, one would expect a decay constant of approximately 11~ns for all scintillator tile sizes, if the photon losses were dominated by absorption in the scintillator.}, but rather by inefficient reflection at the ESR foil or the edge of the tile. 

The rise time, fall time and FWHM of the light collection alone, measured with laser pulses, are smaller than the corresponding numbers inferred for the scintillation process even for the largest scintillator tiles tested. This indicates that, while the tile geometry has an impact on the time resolution, a large contribution comes from the scintillator material itself. In the SiPM-on-tile geometries tested, the time spread of the photons measured at the SiPM is therefore not limited by the geometry but by the scintillator material.

The data points from figure \ref{fig:laser:lc-time-structure} are also listed in table \ref{tab:results-rise-and-fall-times}. For reference, the data sheet values of rise and fall time of BC408 are shown together with the measurements presented in this paper. Since the manufacturer does not state how their time constants are defined, these values are however not directly comparable.

\begin{table}[h!]
  \centering
  \caption{Rise time, fall time and FWHM of photon time distributions for test beam measurements, scintillation and light collection. These values are visualized in figure \ref{fig:laser:lc-time-structure}.}
  \begin{tabular}{l|ccc}
    & Rise time (ns) & Fall time (ns) & FWHM (ns)\\
    \hline
    BC408 time distribution measured & 0.97 $\pm$ 0.07 & 3.36 $\pm$ 0.14 & 2.70 $\pm$ 0.21 \\
    BC408 manufacturer information \cite{BC408_Datasheet} & 0.9 & 2.1 & \\
    \hline
    Light collection \tile{20} & 0.60 $\pm$ 0.04 & 0.55 $\pm$ 0.03 & 0.757 $\pm$ 0.014 \\
    Light collection \tile{30} & 0.71 $\pm$ 0.03 & 1.43 $\pm$ 0.11 & 1.11 $\pm$ 0.03 \\
    Light collection \tile{40} & 0.74 $\pm$ 0.02 & 1.91 $\pm$ 0.08 & 1.22 $\pm$ 0.05 \\
    \hline
    Test Beam \tile{20} & 0.97 $\pm$ 0.09 & 4.26 $\pm$ 0.34 & 4.32 $\pm$ 0.27 \\
    Test Beam \tile{30} & 1.23 $\pm$ 0.09 & 5.3 $\pm$ 0.5   & 5.16 $\pm$ 0.20 \\
    Test Beam \tile{40} & 1.03 $\pm$ 0.09 & 5.7$\pm$ 0.5    & 5.80 $\pm$ 0.30 \\
  \end{tabular}
  \label{tab:results-rise-and-fall-times}
\end{table}

Since the naive scaling of the time resolution with overall light yield, given in equation \ref{eq:energy-binned-tr}, implicitly assumes that the  shape of time distribution of the photons is constant for different amplitudes, this description needs to be refined when considering different tile sizes. The time resolution is determined based on the rising flank of the signal, so only the fraction of the photons in the rising part of the signal, $r_{rise}$, contribute to the time resolution.
We approximate that this fraction is proportional to the inverse of the width of the photon time distribution, $r_{rise} \propto 1/FWHM$. Therefore, equation \ref{eq:energy-binned-tr} is modified by introducing an additional term that corrects for the geometry dependence of the fraction of the signal contributing to the time measurement, given by the inverse of the width of the photon time distribution, resulting in
\begin{equation}
    \sigma_t \propto \frac{1}{\sqrt{n_{photons} \cdot r_{rise}}} \propto \sqrt{\frac{FWHM}{LY}} \propto A^{1/4} \cdot FWHM^{1/2}.
    \label{eq:tr-scale-A-and-FWHM}
\end{equation}

It is thus expected that the quantity $\sigma_{1MIP} \cdot A^{-1/4} \cdot FWHM^{-1/2}$ is constant for different tile sizes. The calculated values, shown in table \ref{tab:results-constant-quantity}, agree with this hypothesis within $2\sigma$, significantly better than with a simple scaling shown in table \ref{tab:tb:tr} which only takes into account the amplitude alone. 

\begin{table}[h!]
  \centering
  \caption{Comparison of the scaling of the time resolution with the tile area and FWHM of the photon time distribution. The right column is expected to be constant for all tile sizes, if the the approximation in equation \ref{eq:tr-scale-A-and-FWHM} holds.}
  \begin{tabular}{l|c}
    Tile size & $\sigma_{1MIP} \cdot A^{-1/4} \cdot FWHM^{-1/2}$ ($\sqrt{\text{ns}/\text{cm}}$) \\
    \hline
    Test Beam \tile{20} & 0.131 $\pm$ 0.008 \\
    Test Beam \tile{30} & 0.147 $\pm$ 0.009 \\
    Test Beam \tile{40} & 0.15 $\pm$ 0.01 \\
  \end{tabular}
  \label{tab:results-constant-quantity}
\end{table}

\section{Conclusions}

Using DESY's test beam facility, we have measured the light yields and time resolutions of SiPM-on-tile configurations with three different scintillator tile sizes, all made of the plastic scintillator BC408 with a thickness of 3~mm. The time resolution for an energy deposition corresponding to minimum ionization is determined with a numerical fit to the energy-dependent time resolution. In addition to the fit uncertainty we include a 6\% fluctuation for the final value of the time resolution that we expect to arise from individual tile-to-tile fluctuations. The measured time resolutions are 
\begin{itemize}
    \item $\sigma_t = 0.38 \pm 0.02$ for \tile{20} scintillator tiles,
    \item $\sigma_t = 0.58 \pm 0.04$ for \tile{30}, and
    \item $\sigma_t = 0.70 \pm 0.04$ for \tile{40}.
\end{itemize}
for an energy deposition corresponding to the most probable value of a minimum-ionizing particle. The time resolution scales with the inverse of the square root of the signal amplitude. This means that the time resolution strongly depends on the energy deposition in the scintillator tile, and thus also on the light yield. The light yield in turn depends on the scintillator tile size.

In order to disentangle additional contributions of scintillation, light collection and light detection to the timing properties, we have developed different measurement methods where the signals are created using electrons from a radioactive $^{90}$Sr source or pulsed laser light.

These complementary measurements show that our SiPM, electronics and digitisation are fast enough to have no impact on our measurement results. Instead, the time structure observed in a SiPM-on-tile configuration can be fully described in terms of the scintillation and the light propagation in the scintillator tile.

The light collection process shows a clear dependence on the scintillator tile size, with larger scintillator tiles delivering broader photon time distributions. While this contribution is small compared to the emission time of the scintillator material itself, it does result in an additional tile-size dependence of the time resolution beyond the effect of the light yield alone, with further improved time resolution for smaller scintillator tiles. 

In order to quantify how the scintillator tile size, the light yield, and the time resolution are correlated, we have developed a Geant4-based simulation of a SiPM-on-Tile detector which will be presented in an upcoming publication.

\section*{Acknowledgements}
We thank our colleagues in the CALICE Collaboration for helpful discussions. We also express our gratitude to the anonymous referee for the excellent suggestions during the review process, which led to substantial improvements of this paper.

This project has received funding from the European Union’s Horizon 2020 Research and Innovation programme under Grant Agreements no. 654168 and no. 101004761.

The measurements leading to these results have been performed at the Test Beam Facility at DESY Hamburg (Germany), a member of the Helmholtz Association (HGF).

\bibliographystyle{JHEP}
\bibliography{STS-Refs}

\end{document}